\documentclass[12pt]{article}

\usepackage[dvipdfmx]{graphicx}
\usepackage{subcaption}
\usepackage{listings,jvlisting}
\usepackage{amsmath,amssymb}
\usepackage{bm}
\usepackage{graphicx, color}
\usepackage{wrapfig}
\usepackage{cite}
\usepackage{braket}
\usepackage{hyperref}
\DeclareMathOperator{\Tr}{Tr}
\begin{document}
\title{
\begin{flushright}
\ \\*[-80pt]
\begin{minipage}{0.2\linewidth}
\normalsize
HUPD-2215 \\*[50pt]
\end{minipage}
\end{flushright}
{\Large \bf
Mass Relations, Unification, and Proton Decay \\
in the Mirror GUT Model
}\\*[20pt]}

\author{
\centerline{
Yusuke~Shimizu $^{1,2}$\footnote{yu-shimizu@hiroshima-u.ac.jp} and
~Shonosuke~Takeshita $^{1}$\footnote{shonosuke@hiroshima-u.ac.jp}}
\\*[20pt]
\centerline{
\begin{minipage}{\linewidth}
\begin{center}
$^1${\it \normalsize
Physics Program, Graduate School of Advanced Science \\ and Engineering,~Hiroshima~University, \\
Higashi-Hiroshima~739-8526,~Japan \\*[5pt]
$^2${\it \normalsize
Core of Research for the Energetic Universe, Hiroshima University, \\
Higashi-Hiroshima 739-8526, Japan}
}
\end{center}
\end{minipage}}
\\*[50pt]}

\date{
\centerline{\small \bf Abstract}
\begin{minipage}{0.9\linewidth}
\medskip
\medskip
\small
We propose an SU(5)$\times $U(1)$_\text{X}\times$U(1)$_\text{PQ}$ model without SUSY because the SUSY particles have not been observed yet. 
The $\mathrm{U(1)_{X}}$ gauge symmetry is the generalization of the $\mathrm{U(1)_{B-L}}$ gauge symmetry and $\mathrm{U(1)_{PQ}}$ symmetry is the global Peccei-Quinn symmetry.
We introduce three mirror families in order to unify the SM gauge couplings and avoid the restriction of proton lifetime. 
In order to obtain the difference of the masses for the down-type quarks and charged leptons, we also introduce the $\mathbf{45}$ representation Higgs in addition to the $\mathbf{5}$ representation one.
In this paper, we discuss the mass relations between the SM and mirror particles and identify the mass scales of the mirror particles.
Since the new particles exist in the intermediate energy scale and contribute to the renormalization group equation, the SM gauge couplings unify successfully at high energy.
Our model can be tested by the future proton decay search, e.g., the Hyper-Kamiokande experiment expected as $\tau_p(p\to\pi^0{e^+})<1.0\times10^{35}$ years.
Also, by using the mass relations between the active and mirror neutrinos, we estimate the lower bound of the heavy neutrino masses.
\end{minipage}
}

\begin{titlepage}
\maketitle
\thispagestyle{empty}
\end{titlepage}
\newpage
\section{Introduction}
The standard model (SM) is the successful one which includes $\mathrm{SU(3)_C}$, $\mathrm{SU(2)_L}$, 
and $\mathrm{U(1)_Y}$ gauge symmetries. 
The SM has been completed with the discovery of the Higgs boson in 2012~\cite{ATLAS:2012yve, CMS:2012qbp}.
However, the SM cannot be explained 
the origins of the charges and flavor structures for the SM particles and so on. 
Then, we need new physics beyond the SM to solve these problems. 
One of the attractive theories is the grand unified theory (GUT). 
The main feature of the GUT is the unification of strong, weak, and electromagnetic forces.
The minimal model of the GUT is the SU(5) one~\cite{Georgi:1974sy}. 
In the minimal SU(5) model, the SM matter fields and gauge bosons are unified into ${\bf \bar 5}$, ${\bf 10}$, and ${\bf 24}$ adjoint representations, respectively.
Thanks to introducing the $\mathbf{45}$ representation Higgs field~\cite{Kalyniak:1982pt, Eckert:1983bn} in this model, we obtain the difference of the masses for the down-type quarks and charged leptons~\cite{Georgi:1979df}.  
In addition to the unification of the SM particles and gauge interactions, the GUT predicts proton decay. 
Then, the GUT can be tested by the proton decay search.
However, there are several problems with the GUT. 
First, the proton lifetime expected by the GUT is inconsistent with the current proton decay search results, for example, the Super-Kamiokande experiment reported $\tau_p(p\to\pi^0{e^+})>2.4\times10^{34}$ years~\cite{Super-Kamiokande:2020wjk}.
Especially, the minimal SU(5) GUT model expects the proton lifetime such as $\tau_p(p\to\pi^0{e^+})\approx10^{30}$-$10^{31}$ years~\cite{Georgi:1974yf}.
On the other hand, by solving the renormalization group equations (RGE), we can get the SM gauge couplings at the high energy scale. 
However, the SM gauge couplings do not unify successfully at high energy.
This is another problem with the GUT.
Considering the supersymmetry (SUSY) in the theory, the SM gauge couplings unify successfully at high energy~\cite{Dimopoulos:1981yj, Marciano:1981un, Einhorn:1981sx, Amaldi:1991cn, Langacker:1991an, Ellis:1990wk, Giunti:1991ta}.
Since the SUSY particles have not been observed yet in the experiments, we need some extensions of the GUT model except for SUSY. 

In this paper, we propose an SU(5)$\times $U(1)$_\text{X}\times$U(1)$_\text{PQ}$ model without SUSY. 
The $\mathrm{U(1)_{X}}$ gauge symmetry~\cite{Appelquist:2002mw} is the generalization of the $\mathrm{U(1)_{B-L}}$ (baryon minas lepton number) gauge symmetry~\cite{Davidson:1978pm, Mohapatra:1980qe, Marshak:1979fm, Wetterich:1981bx, Masiero:1982fi, Mohapatra:1982xz, Buchmuller:1991ce}.
The definition of the $\mathrm{U(1)_{X}}$ charge is $Q_\mathrm{X}=x_H~Q_\mathrm{Y}+Q_\mathrm{B-L}$, where $Q_\mathrm{Y}$ is hypercharge, $Q_\mathrm{B-L}$ is B-L charge, and $x_H$ is a free parameter~\cite{Oda:2015gna}.
For $x_H=-4/5$~\cite{Okada:2017dqs}, the $\mathrm{U(1)_{X}}$ charge can be assigned to the ${\bf \bar 5}$ and ${\bf 10}$ representations successfully.
The three gauge singlet right-handed Majorana neutrinos cancel all the $\mathrm{U(1)_{X}}$ related anomalies.
The $\mathrm{U(1)_{PQ}}$ symmetry is the global Peccei-Quinn (PQ) symmetry~\cite{Peccei:1977hh, Peccei:1977ur}.
The PQ symmetry can solve the strong CP problem~\cite{Peccei:2006as} and the pseudo-Nambu Goldstone boson from the PQ symmetry breaking called the axion is the dark matter (DM) candidate~\cite{Weinberg:1977ma, Wilczek:1977pj}.
In previous work~\cite{Okada:2021skz}, they added one family of the same representation of the SM particles and one mirror family.
The mirror family is the conjugate of the representation for the SM particles.
In our model, we introduce only three mirror families~\cite{Blinnikov:1982eh, Blinnikov:1983gh, Khlopov:1989fj} in order to unify the SM gauge couplings successfully and avoid the restriction of proton lifetime.
In order to obtain the difference of the masses for the down-type quarks and charged leptons,
we also introduce the $\mathbf{45}$ representation Higgs in addition to the $\mathbf{5}$ representation one. 
Then, we discuss the mass relations between the SM and mirror particles and identify the mass scales of the mirror particles.
We set the benchmark for the mass scales of the mirror particles in our analysis.
Since the new particles exist in the intermediate energy scale and contribute to the RGE, the SM gauge couplings unify successfully at $M_\mathrm{GUT}\approx{7.28}\times10^{15}$~GeV and $\alpha_\mathrm{GUT}=\alpha_1=\alpha_2=\alpha_3\approx{1/31.7}$. 
Our model expects the proton lifetime as $\tau_p(p\to\pi^0{e^+})\approx{8.07\times10^{34}}~\mathrm{years}$ and can be tested by the future proton decay search, for example, the Hyper-Kamiokande experiment expected as $\tau_p(p\to\pi^0{e^+})<1.0\times10^{35}$ years~\cite{Dealtry:2019ldr}.
Also, by using the mass relations between the active and mirror neutrinos, we estimate the lower bound of the heavy neutrino masses.
The lowest bound of the heavy neutrino masses is ${m_{\mathrm{heavy}}}>104.34~\mathrm{GeV}$ and it is testable for the future experiment about the sterile neutrinos~\cite{Alekhin:2015byh}.

This paper is organized as follows. In section~\ref{sec:Model}, we propose an 
SU(5)$\times $U(1)$_X\times$U(1)$_\text{PQ}$ model and discuss the scenario of the spontaneous symmetry breaking with 
vacuum expectation values (VEVs) of the scalar fields.
We also discuss the Yukawa couplings for the SM and 
mirror particles after spontaneous symmetry breaking. 
In section~\ref{sec:mass_relation}, we derive the mass relations between the SM and mirror particles. 
Section~\ref{sec:proton_decay} is shown the identification of the new particles, the SM gauge unification, and proton decay in our model.
In section~\ref{sec:neutrino_mass}, we obtain the mass relations between the active and mirror neutrinos and estimate the heavy neutrino masses. 
Section~\ref{sec:Summary} is devoted to summary. 
In appendix~\ref{sec:Representations}, we show the unification of the SM particles into the SU(5) representations.
Appendix~\ref{sec:beta_coefficients} is shown the RGE form and the beta coefficients of the relevant particles. 

\label{sec:Intro}
\section{SU(5)$\times $U(1)$_\text{X}\times$U(1)$_\text{PQ}$ model}
\label{sec:Model}
In this section, we propose an SU(5)$\times $U(1)$_\text{X}\times$U(1)$_\text{PQ}$ model without SUSY. 
We introduce two symmetries, the $\mathrm{U(1)_{X}}$ gauge symmetry~\cite{Appelquist:2002mw} and $\mathrm{U(1)_{PQ}}$ symmetry~\cite{Peccei:1977hh, Peccei:1977ur}, in addition to the SU(5) symmetry.
The $\mathrm{U(1)_{X}}$ gauge symmetry is the generalization of the $\mathrm{U(1)_{B-L}}$ gauge symmetry~\cite{Davidson:1978pm, Mohapatra:1980qe, Marshak:1979fm, Wetterich:1981bx, Masiero:1982fi, Mohapatra:1982xz, Buchmuller:1991ce}.
The definition of the $\mathrm{U(1)_{X}}$ charge is $Q_\mathrm{X}=x_H~Q_\mathrm{Y}+Q_\mathrm{B-L}$, where $Q_\mathrm{Y}$ is hypercharge, $Q_\mathrm{B-L}$ is B-L charge, and $x_H$ is a free parameter~\cite{Oda:2015gna}.
For $x_H=-4/5$~\cite{Okada:2017dqs}, the $\mathrm{U(1)_{X}}$ charge can be assigned to the ${\bf \bar 5}$ and ${\bf 10}$ representations successfully.
The $\mathrm{U(1)_{PQ}}$ symmetry is the global Peccei-Quinn (PQ) symmetry.
The PQ symmetry can solve the strong CP problem~\cite{Peccei:2006as} and the axion which is the pseudo-Nambu Goldstone boson from the PQ symmetry breaking is the dark matter (DM) candidate~\cite{Weinberg:1977ma, Wilczek:1977pj}.
In the SU(5) model, the SM matter fields are unified into a 
$\mathbf{\bar 5}$ and $\mathbf{10}$ representations defined as $\psi _{\bar 5}^i$ and 
$\psi _{10}^i$, respectively ($i=1$-$3$ generations).
In appendix~\ref{sec:Representations}, we describe the unification of the SM particles into the SU(5) representations in detail.
In order to unify the SM gauge couplings successfully and avoid the restriction of proton lifetime, 
we introduce three mirror families which are defined as 
$\Tilde{\psi}_5^i$ and $\Tilde{\psi}_{\overline{10}}^i$ ($i=1$-$3$) for $\mathbf{5}$ and $\mathbf{\overline{10}}$ representations in SU(5), respectively.
The mirror family is the conjugate of the representation for the SM particles.
In our model, three mirror families have the $\mathrm{U(1)_{PQ}}$ charge~\cite{Davidson:1981zd, Davidson:1983fy, Davidson:1983fe}.
We also introduce three gauge singlet right-handed Majorana neutrinos $(N^c)^i$ ($i=1$-$3$) in order to derive the left-handed Majorana neutrino masses through the type-I seesaw mechanism~\cite{Minkowski:1977sc, Yanagida, Gell-Mann:1979vob, Mohapatra:1979ia, Schechter:1980gr} and cancel all the U(1)$_\text{X}$ related anomalies.
Here, $``c"$ means charge conjugation.
In addition to the $\mathbf{5}$ representation Higgs particle which is defined as $H$, we introduce $\mathbf{24}$ and $\mathbf{45}$ representation Higgs fields~\cite{Kalyniak:1982pt, Eckert:1983bn} defined as $\Sigma $ and $\chi $, respectively. 
The $\chi$ satisfies the following conditions as $\chi^{ab}_{c}=-\chi^{ba}_{c},~{\chi}^{ab}_{a}=0~(a,b,c=1$-$5)$.
Thanks to the $\mathbf{45}$ representation Higgs field $\chi $, we obtain the difference of the masses for the down-type quarks and charged leptons~\cite{Georgi:1979df}. 
The gauge singlet scalar field $\Phi$ is introduced in order to occur inflation and break the U(1)$_\text{X}$ gauge symmetry. 
The assignments of the particles for SU(5)$\times $U(1)$_\text{X}\times$U(1)$_\text{PQ}$ symmetry are summarized in Table~\ref{tab:model}.

\begin{table}[h]
    \centering
    \begin{tabular}{c|c|c|c}
    &SU(5)&$\mathrm{U(1)_X}$&$\mathrm{U(1)_{PQ}}$\\
    \hline \rule[14pt]{0pt}{0pt}
        $\psi_{\bar{5}}^i$&$\mathbf{\bar{5}}$&$-3/5$&0\\
     \hline \rule[14pt]{0pt}{0pt}
        $\psi_{10}^i$&$\mathbf{10}$&$1/5$&0\\
    \hline \rule[14pt]{0pt}{0pt}
        $\Tilde{\psi}_{5}^i$&$\mathbf{5}$&$3/5$&1\\
     \hline \rule[14pt]{0pt}{0pt}
     $\Tilde{\psi}_{\overline{10}}^i$&$\mathbf{\overline{10}}$&$-1/5$&1\\
     \hline \rule[14pt]{0pt}{0pt}
     $(N^c)^i$&$\mathbf{1}$&+1&0\\
     \hline \rule[14pt]{0pt}{0pt}
     $\Sigma$&$\mathbf{24}$&0&$-1$\\
     \hline \rule[14pt]{0pt}{0pt}
     $\chi$&$\mathbf{45}$&$-2/5$&0\\
     \hline \rule[14pt]{0pt}{0pt}
     $\Phi$&$\mathbf{1}$&$-2$&0\\
     \hline \rule[14pt]{0pt}{0pt}
     $H$&$\mathbf{5}$&$-2/5$&0\\
    \end{tabular}
    \caption{The assignments of the particles for SU(5)$\times $U(1)$_\text{X}\times$U(1)$_\text{PQ}$ symmetry.}
    \label{tab:model}
\end{table}

In our model, the scenario of spontaneous symmetry breaking is following three steps. At first, the $\mathbf{24}$ representation 
Higgs $\Sigma $ breaks SU(5) and U(1)$_\text{PQ}$ symmetries by taking the VEVs as 
$\langle\Sigma\rangle=v_{\Sigma}/(2\sqrt{15})\mathrm{Diag}\left (-2,-2,-2,3,3\right )$. 
Next, the gauge singlet scalar field $\Phi $ breaks U(1)$_\text{X}$ symmetry after taking VEV as 
$\langle\Phi\rangle=v_\Phi/\sqrt{2}$. Finally, the $\mathbf{5}$ and $\mathbf{45}$ representation Higgs, $H$ and $\chi$, break the SM gauge symmetries by taking VEVs as $\langle H \rangle=\left (0~0~0~0~v_H/\sqrt{2}\right )^T$, 
$\langle{\chi}\rangle^{15}_{1}=\langle{\chi}\rangle^{25}_{2}=\langle{\chi}\rangle^{35}_{3}=v_{\chi}/\sqrt{2}$, and
$\langle{\chi}\rangle^{45}_{4}=-3v_{\chi}/\sqrt{2}$, respectively.

Let us discuss the Yukawa interactions which are invariant under the SU(5)$\times $U(1)$_\text{X}\times$U(1)$_\text{PQ}$ symmetry. 
In our model, we can write the following Lagrangian which includes three types of Yukawa interactions:
\begin{equation}
\mathcal{L}_Y\supset\mathcal{L}_{\mathrm{SM}}+\mathcal{L}_{\mathrm{mirror}}+\mathcal{L}_{\mathrm{neutrino}}~,
\label{eq:Yukawa_Lagrangian}
\end{equation}
where the relevant SM particles are written as 
\begin{equation}
\mathcal{L}_{\mathrm{SM}}\supset\sum_{i,j=1}^{3}[(Y^{ij}_{1}H+Y^{ij}_{2}\chi)\psi_{10}^i\psi_{10}^j]+\sum_{i,j=1}^{3}[(Y^{ij}_{3}H^*+Y^{ij}_{4}\chi)\psi_{\bar{5}}^i\psi_{10}^j],
\label{eq:SM_Lagrangian}
\end{equation}
and the Lagrangian for the SM and mirror particles is obtained as
\begin{equation}
\mathcal{L}_{\mathrm{mirror}}\supset\sum_{i,j=1}^{3}\Tilde{Y}_{5}^{ij}\Sigma\psi_{\bar{5}}^i\Tilde{\psi}_{5}^j+\Tr\left[\sum_{i,j=1}^{3}\Tilde{Y}_{10}^{ij}\Sigma\psi_{10}^i\Tilde{\psi}_{\overline{10}}^j\right],
\label{eq:mirror_Lagrangian}
\end{equation}
and the Lagrangian for the Dirac neutrinos and right-handed Majorana neutrinos is given as
\begin{equation}
\mathcal{L}_{\mathrm{neutrino}}\supset-\sum_{i,j=1}^{3}Y_D^{ij}H\psi_{\bar{5}}^i(N^c)^j-\left(\frac{1}{2}\sum_{i=1}^{3}Y_M^i\Phi(N^c)^i{(N^c)^i}+\mathrm{h.c.}\right).
\label{eq:neutrino_Lagrangian}
\end{equation}
After spontaneous symmetry breaking, we obtain the mass matrices 
from Eq.~\eqref{eq:Yukawa_Lagrangian} as follows:
\begin{align}
    \mathcal{M}_U&=
    \begin{pmatrix}
    2\sqrt{2}Y^{ij}_{1}{v_H}&-\Tilde{Y}_{10}^{ij}\frac{v_\Sigma}{4\sqrt{15}}\\
    \Tilde{Y}_{10}^{ij}\frac{4v_\Sigma}{\sqrt{15}}&0\\
    \end{pmatrix}, \nonumber \\
    \mathcal{M}_D&=
    \begin{pmatrix}
    \frac{1}{2}[Y^{ji}_{3}{v_H^*}+2Y^{ji}_4v^*_{\chi}]&-\Tilde{Y}_{10}^{ij}\frac{v_\Sigma}{4\sqrt{15}}\\
    -\Tilde{Y}_{5}^{ij}\frac{v_\Sigma}{2\sqrt{15}}&0\\
    \end{pmatrix}, \label{eq:mass_matrices} \\
    \mathcal{M}_E&=
    \begin{pmatrix}
    \frac{1}{2}[Y^{ij}_{3}{v_H^*}-6Y^{ij}_4v^*_{\chi}]&\Tilde{Y}_{5}^{ij}\frac{3v_\Sigma}{4\sqrt{15}}\\
    -\Tilde{Y}_{10}^{ij}\frac{3v_\Sigma}{2\sqrt{15}}&0\\
    \end{pmatrix}, \nonumber \\
    \mathcal{M}_\nu&=
    \begin{pmatrix}
        0&\Tilde{M}_D&M_D\\
        \Tilde{M}_D^T&0&0\\
        M_D^T&0&M_R\\
    \end{pmatrix}\nonumber ,
    \end{align}
where $\mathcal{M}_U$, $\mathcal{M}_D$, and $\mathcal{M}_E$ are the $6\times 6$ mass matrices which include the SM and mirror particles 
for the up- and down-type quark and charged lepton sectors, respectively.
On the other hand, $\mathcal{M}_\nu $ is the $9\times 9$ mass matrix which includes 
the Dirac, mirror, and right-handed Majorana neutrinos. 
By using Eq.~\eqref{eq:mass_matrices}, we obtain the mass eigenvalues and derive the mass relations between the SM and mirror particles in the next section. Note that mass relations among neutrinos are shown 
in section~\ref{sec:neutrino_mass}.
%
%

\section{Mass relations for the SM and mirror particles}
\label{sec:mass_relation}
In the previous section, we have discussed the Yukawa interactions for the SM and mirror particles and have obtained the $6\times 6$ mass matrices for the up- and down-type quarks and charged leptons. 
Let us discuss the mass relations for the SM and mirror particles in this section.

First, we obtain the mass eigenvalues for the up- and down-type quark and charged lepton sectors.
In our analysis, we assume the diagonal matrices for each particle for simplicity.
By using the $6\times6$ mass matrices $\mathcal{M}_U$ in Eq.~\eqref{eq:mass_matrices}, we analyze the up-type quark mass eigenvalues $m_{ui}$~($i=1$-$6$).
The up-type quark squared mass eigenvalues are written just below:
\begin{align}
        m^2_{u1}=\frac{1}{480}&\Bigg[17{\big(v_\Sigma\Tilde{Y}_{10}^{11}\big)}^2+1920{\big(v_{H}Y_{1}^{11}\big)}^2\nonumber\\
&-\sqrt{-64{\big(v_\Sigma\Tilde{Y}_{10}^{11}\big)}^4+\Big(17{\big(v_\Sigma\Tilde{Y}_{10}^{11}\big)}^2+1920{\big(v_{H}Y_{1}^{11}\big)}^2\Big)^2}\Bigg],\nonumber\\
        m^2_{u2}=\frac{1}{480}&\Bigg[17{\big(v_\Sigma\Tilde{Y}_{10}^{11}\big)}^2+1920{\big(v_{H}Y_{1}^{11}\big)}^2\nonumber\\
&+\sqrt{-64{\big(v_\Sigma\Tilde{Y}_{10}^{11}\big)}^4+\Big(17{\big(v_\Sigma\Tilde{Y}_{10}^{11}\big)}^2+1920{\big(v_{H}Y_{1}^{11}\big)}^2\Big)^2}\Bigg],\nonumber\\
        m^2_{u3}=\frac{1}{480}&\Bigg[17{\big(v_\Sigma\Tilde{Y}_{10}^{22}\big)}^2+1920{\big(v_{H}Y_{1}^{22}\big)}^2\nonumber\\
        &-\sqrt{-64{\big(v_\Sigma\Tilde{Y}_{10}^{22}\big)}^4+\Big(17{\big(v_\Sigma\Tilde{Y}_{10}^{22}\big)}^2+1920{\big(v_{H}Y_{1}^{22}\big)}^2\Big)^2}\Bigg],\nonumber\\
        m^2_{u4}=\frac{1}{480}&\Bigg[17{\big(v_\Sigma\Tilde{Y}_{10}^{22}\big)}^2+1920{\big(v_{H}Y_{1}^{22}\big)}^2\label{eq:uptypemasseigenvalues}\\
&+\sqrt{-64{\big(v_\Sigma\Tilde{Y}_{10}^{22}\big)}^4+\Big(17{\big(v_\Sigma\Tilde{Y}_{10}^{22}\big)}^2+1920{\big(v_{H}Y_{1}^{22}\big)}^2\Big)^2}\Bigg],\nonumber\\
        m^2_{u5}=\frac{1}{480}&\Bigg[17{\big(v_\Sigma\Tilde{Y}_{10}^{33}\big)}^2+1920{\big(v_{H}Y_{1}^{33}\big)}^2\nonumber\\
        &-\sqrt{-64{\big(v_\Sigma\Tilde{Y}_{10}^{33}\big)}^4+\Big(17{\big(v_\Sigma\Tilde{Y}_{10}^{33}\big)}^2+1920{\big(v_{H}Y_{1}^{33}\big)}^2\Big)^2}\Bigg],\nonumber\\
        m^2_{u6}=\frac{1}{480}&\Bigg[17{\big(v_\Sigma\Tilde{Y}_{10}^{33}\big)}^2+1920{\big(v_{H}Y_{1}^{33}\big)}^2\nonumber\\
&+\sqrt{-64{\big(v_\Sigma\Tilde{Y}_{10}^{33}\big)}^4+\Big(17{\big(v_\Sigma\Tilde{Y}_{10}^{33}\big)}^2+1920{\big(v_{H}Y_{1}^{33}\big)}^2\Big)^2}\Bigg]\nonumber.
    \end{align}
The up-type quark mass eigenvalues include the $\Tilde{Y}_{10}$ matrix diagonal elements.
Then, the $\Tilde{Y}_{10}$ diagonal elements can be expressed in terms of the up-type quark mass eigenvalues.
On the other hand, the down-type quark and charged lepton mass eigenvalues are related to $\Tilde{Y}_{10}$ matrix diagonal elements.
Therefore, we substitute these values for the down-type quark and charged lepton mass eigenvalues and represent the down-type quark $m_{di}$ and charged lepton mass eigenvalues $m_{ei}$ in terms of the up-type quark mass eigenvalues $(i=1$-$6)$.
We show the down-type quark squared mass eigenvalues just below:
\begin{align}
    m^2_{d1}=&\frac{1}{120}\Bigg[15m_{u1}m_{u2}+15\big(v_{H}Y_{3}^{11}+2v_{\chi}Y_{4}^{11}\big)^2+{\big({v_\Sigma}\Tilde{Y}_5^{11}\big)}^2\nonumber \\
    &-\sqrt{-60m_{u1}m_{u2}{\big({v_\Sigma}\Tilde{Y}_5^{11}\big)}^2+\Big(15m_{u1}m_{u2}+15\big(v_{H}Y_{3}^{11}+2v_{\chi}Y_{4}^{11}\big)^2+{\big({v_\Sigma}\Tilde{Y}_5^{11}\big)}^2\Big)^2}\Bigg],\nonumber \\
    m^2_{d2}=&\frac{1}{120}\Bigg[15m_{u1}m_{u2}+15\big(v_{H}Y_{3}^{11}+2v_{\chi}Y_{4}^{11}\big)^2+{\big({v_\Sigma}\Tilde{Y}_5^{11}\big)}^2\nonumber \\
    &+\sqrt{-60m_{u1}m_{u2}{\big({v_\Sigma}\Tilde{Y}_5^{11}\big)}^2+\Big(15m_{u1}m_{u2}+15\big(v_{H}Y_{3}^{11}+2v_{\chi}Y_{4}^{11}\big)^2+{\big({v_\Sigma}\Tilde{Y}_5^{11}\big)}^2\Big)^2}\Bigg],\nonumber \\
    m^2_{d3}=&\frac{1}{120}\Bigg[15m_{u3}m_{u4}+15\big(v_{H}Y_{3}^{22}+2v_{\chi}Y_{4}^{22}\big)^2+{\big({v_\Sigma}\Tilde{Y}_5^{22}\big)}^2\nonumber \\
    &-\sqrt{-60m_{u3}m_{u4}{\big({v_\Sigma}\Tilde{Y}_5^{22}\big)}^2+\Big(15m_{u3}m_{u4}+15\big(v_{H}Y_{3}^{22}+2v_{\chi}Y_{4}^{22}\big)^2+{\big({v_\Sigma}\Tilde{Y}_5^{22}\big)}^2\Big)^2}\Bigg],\nonumber \\
    m^2_{d4}=&\frac{1}{120}\Bigg[15m_{u3}m_{u4}+15\big(v_{H}Y_{3}^{22}+2v_{\chi}Y_{4}^{22}\big)^2+{\big({v_\Sigma}\Tilde{Y}_5^{22}\big)}^2\label{eq:downtypemasseigenvalues}\\
    &+\sqrt{-60m_{u3}m_{u4}{\big({v_\Sigma}\Tilde{Y}_5^{22}\big)}^2+\Big(15m_{u3}m_{u4}+15\big(v_{H}Y_{3}^{22}+2v_{\chi}Y_{4}^{22}\big)^2+{\big({v_\Sigma}\Tilde{Y}_5^{22}\big)}^2\Big)^2}\Bigg],\nonumber\\
    m^2_{d5}=&\frac{1}{120}\Bigg[15m_{u5}m_{u6}+15\big(v_{H}Y_{3}^{33}+2v_{\chi}Y_{4}^{33}\big)^2+{\big({v_\Sigma}\Tilde{Y}_5^{33}\big)}^2\nonumber \\
    &-\sqrt{-60m_{u5}m_{u6}{\big({v_\Sigma}\Tilde{Y}_5^{33}\big)}^2+\Big(15m_{u5}m_{u6}+15\big(v_{H}Y_{3}^{33}+2v_{\chi}Y_{4}^{33}\big)^2+{\big({v_\Sigma}\Tilde{Y}_5^{33}\big)}^2\Big)^2}\Bigg],\nonumber \\
    m^2_{d6}=&\frac{1}{120}\Bigg[15m_{u5}m_{u6}+15\big(v_{H}Y_{3}^{33}+2v_{\chi}Y_{4}^{33}\big)^2+{\big({v_\Sigma}\Tilde{Y}_5^{33}\big)}^2\nonumber \\
    &+\sqrt{-60m_{u5}m_{u6}{\big({v_\Sigma}\Tilde{Y}_5^{33}\big)}^2+\Big(15m_{u5}m_{u6}+15\big(v_{H}Y_{3}^{33}+2v_{\chi}Y_{4}^{33}\big)^2+{\big({v_\Sigma}\Tilde{Y}_5^{33}\big)}^2\Big)^2}\Bigg],\nonumber
    \end{align}
and the charged lepton squared mass eigenvalues are written as    
    \begin{align}
m^2_{e1}=&\frac{1}{160}\Bigg[720m_{u1}m_{u2}+20\big(v_{H}Y_{3}^{11}-6v_{\chi}Y_{4}^{11}\big)^2+3{\big({v_\Sigma}\Tilde{Y}_5^{11}\big)}^2\nonumber \\
    &-\sqrt{-8640m_{u1}m_{u2}{({v_\Sigma}\Tilde{Y}_5^{11})}^2+\Big(720m_{u1}m_{u2}+20\big(v_{H}Y_{3}^{11}-6v_{\chi}Y_{4}^{11}\big)^2+{3\big({v_\Sigma}\Tilde{Y}_5^{11}\big)}^2\Big)^2}\Bigg],\nonumber \\
    m^2_{e2}=&\frac{1}{160}\Bigg[720m_{u1}m_{u2}+20\big(v_{H}Y_{3}^{11}-6v_{\chi}Y_{4}^{11}\big)^2+3{\big({v_\Sigma}\Tilde{Y}_5^{11}\big)}^2\nonumber \\
    &+\sqrt{-8640m_{u1}m_{u2}{({v_\Sigma}\Tilde{Y}_5^{11})}^2+\Big(720m_{u1}m_{u2}+20\big(v_{H}Y_{3}^{11}-6v_{\chi}Y_{4}^{11}\big)^2+{3\big({v_\Sigma}\Tilde{Y}_5^{11}\big)}^2\Big)^2}\Bigg],\nonumber \\
    m^2_{e3}=&\frac{1}{160}\Bigg[720m_{u3}m_{u4}+20\big(v_{H}Y_{3}^{22}-6v_{\chi}Y_{4}^{22}\big)^2+3{\big({v_\Sigma}\Tilde{Y}_5^{22}\big)}^2\nonumber \\
    &-\sqrt{-8640m_{u3}m_{u4}{({v_\Sigma}\Tilde{Y}_5^{22})}^2+\Big(720m_{u3}m_{u4}+20\big(v_{H}Y_{3}^{22}-6v_{\chi}Y_{4}^{22}\big)^2+{3\big({v_\Sigma}\Tilde{Y}_5^{22}\big)}^2\Big)^2}\Bigg],\nonumber \\
    m^2_{e4}=&\frac{1}{160}\Bigg[720m_{u3}m_{u4}+20\big(v_{H}Y_{3}^{22}-6v_{\chi}Y_{4}^{22}\big)^2+3{\big({v_\Sigma}\Tilde{Y}_5^{22}\big)}^2\label{eq:chargedleptonmasseigenvalues}\\
    &+\sqrt{-8640m_{u3}m_{u4}{({v_\Sigma}\Tilde{Y}_5^{22})}^2+\Big(720m_{u3}m_{u4}+20\big(v_{H}Y_{3}^{22}-6v_{\chi}Y_{4}^{22}\big)^2+{3\big({v_\Sigma}\Tilde{Y}_5^{22}\big)}^2\Big)^2}\Bigg],\nonumber \\
    m^2_{e5}=&\frac{1}{160}\Bigg[720m_{u5}m_{u6}+20\big(v_{H}Y_{3}^{33}-6v_{\chi}Y_{4}^{33}\big)^2+3{\big({v_\Sigma}\Tilde{Y}_5^{33}\big)}^2\nonumber \\
    &-\sqrt{-8640m_{u5}m_{u6}{({v_\Sigma}\Tilde{Y}_5^{33})}^2+\Big(720m_{u5}m_{u6}+20\big(v_{H}Y_{3}^{33}-6v_{\chi}Y_{4}^{33}\big)^2+{3\big({v_\Sigma}\Tilde{Y}_5^{33}\big)}^2\Big)^2}\Bigg],\nonumber \\
    m^2_{e6}=&\frac{1}{160}\Bigg[720m_{u5}m_{u6}+20\big(v_{H}Y_{3}^{33}-6v_{\chi}Y_{4}^{33}\big)^2+3{\big({v_\Sigma}\Tilde{Y}_5^{33}\big)}^2\nonumber \\
    &+\sqrt{-8640m_{u5}m_{u6}{({v_\Sigma}\Tilde{Y}_5^{33})}^2+\Big(720m_{u5}m_{u6}+20\big(v_{H}Y_{3}^{33}-6v_{\chi}Y_{4}^{33}\big)^2+{3\big({v_\Sigma}\Tilde{Y}_5^{33}\big)}^2\Big)^2}\Bigg].\nonumber 
\end{align}

Now we obtain the mass eigenvalues for the up- and down-type quark and charged lepton sectors.
Next, by using these mass eigenvalues, we derive the mass relations for the SM and mirror particles.
The down-type quark and charged lepton mass eigenvalues are represented in terms of the up-type quark mass eigenvalues.
Then, we can derive the mass relations between the up-type quark and the down-type quark mass eigenvalues and between the up-type quark and the charged lepton mass eigenvalues,
\begin{equation}
\label{eq:massrelation}
\begin{aligned}
m^2_{d1}m^2_{d2}&=\frac{m_{u1}m_{u2}\big({v_\Sigma}\Tilde{Y}_5^{11}\big)^2}{240},\quad{m^2_{e1}m^2_{e2}}=\frac{27m_{u1}m_{u2}\big({v_\Sigma}\Tilde{Y}_5^{11}\big)^2}{80},\\
m^2_{d3}m^2_{d4}&=\frac{m_{u3}m_{u4}\big({v_\Sigma}\Tilde{Y}_5^{22}\big)^2}{240},\quad{m^2_{e3}m^2_{e4}}=\frac{27m_{u3}m_{u4}\big({v_\Sigma}\Tilde{Y}_5^{22}\big)^2}{80},\\
m^2_{d5}m^2_{d6}&=\frac{m_{u5}m_{u6}\big({v_\Sigma}\Tilde{Y}_5^{33}\big)^2}{240},\quad{m^2_{e5}m^2_{e6}}=\frac{27m_{u5}m_{u6}\big({v_\Sigma}\Tilde{Y}_5^{33}\big)^2}{80}.\\
\end{aligned}
\end{equation}
By using these mass relations, we identify the mass scale of the mirror particles in the next section.

\section{Gauge unification and proton decay}
\label{sec:proton_decay}
In this section, we discuss the contribution of the new particles to the RGE and proton lifetime.
In order to obtain the RGE including the contribution of the new particles, we identify the mass scales of the new particles by using the mass relations in Eq.~\eqref{eq:massrelation}.
Once, we assume that $m_{d1}$ is bottom quark (4.18~GeV) and $m_{e1}$ is electron (0.511~MeV)~\cite{ParticleDataGroup:2022pth}. 
Then, from the mass relations in Eq.~\eqref{eq:massrelation}, we can identify $m_{d2}=\mathcal{O}(10^3)$~GeV, $m_{e2}=7.36\times{10}^{4}\times{m_{d2}}$~GeV.
The others are the SM or $\mathcal{O}(M_\mathrm{GUT})$ scale particles that satisfied the mass relations.
We show the fermion mass setup in Table~\ref{table:fermionmasssetup}.
\begin{table}[h]
    \centering
    \begin{tabular}{|c|c||c|c||c|c|}
         \hline
    $m_{u1}$&SM particle&$m_{d1}$&bottom quark&$m_{e1}$&electron\\
         \hline
$m_{u2}$&$\mathcal{O}(M_\mathrm{GUT})$&$m_{d2}$&$\mathcal{O}(10^3)$~GeV&$m_{e2}$&$7.36\times10^4\times{m_{d2}}$~GeV\\
         \hline
$m_{u3}$&SM particle&$m_{d3}$&SM particle&$m_{e3}$&SM particle\\
         \hline
$m_{u4}$&$\mathcal{O}(M_\mathrm{GUT})$&$m_{d4}$&$\mathcal{O}(M_\mathrm{GUT})$&$m_{e4}$&$\mathcal{O}(M_\mathrm{GUT})$\\
         \hline
$m_{u5}$&SM particle&$m_{d5}$&SM particle&$m_{e5}$&SM particle\\
         \hline
$m_{u6}$&$\mathcal{O}(M_\mathrm{GUT})$&$m_{d6}$&$\mathcal{O}(M_\mathrm{GUT})$&$m_{e6}$&$\mathcal{O}(M_\mathrm{GUT})$\\
         \hline
    \end{tabular}
    \caption{The mass setup of the SM and mirror particles.}
    \label{table:fermionmasssetup}
\end{table}

In addition to the fermion masses, we identify the scalar mass scale in order to unify the SM gauge couplings.
The $\mathbf{45}$ representation Higgs $\chi$ is composed of
\begin{align}
    \chi\sim\Phi_1(8,2,\frac{1}{2})\oplus\Phi_2(\bar{6},1,-\frac{1}{3})&\oplus\Phi_3(3,3,-\frac{1}{3})\nonumber\\
    &\oplus\Phi_4(\bar{3},2,-\frac{7}{6})
\oplus\Phi_5(3,1,-\frac{1}{3})\oplus\Phi_6(\bar{3},1,\frac{4}{3})\oplus{H_2}(1,2,\frac{1}{2}). 
\end{align}
We assume that the scalar $\phi_1$ mass is $M_1=\mathcal{O}(10^3)$~GeV, the scalar $\phi_3$ mass is $M_3=\mathcal{O}(10^9)$~GeV, and the remained scalar masses are $\mathcal{O}(M_\mathrm{GUT})$ scale.
Therefore, four new particles, $m_{d2}$, $m_{e2}$, $\phi_1$, and $\phi_3$, exist in the intermediate energy scale and contribute to the achieving of unifying the SM gauge couplings successfully.

\begin{figure}[htbp]
\begin{tabular}{cc}
    \begin{minipage}[t]{0.45\hsize}
    \centering
    \includegraphics[keepaspectratio,scale=0.45]{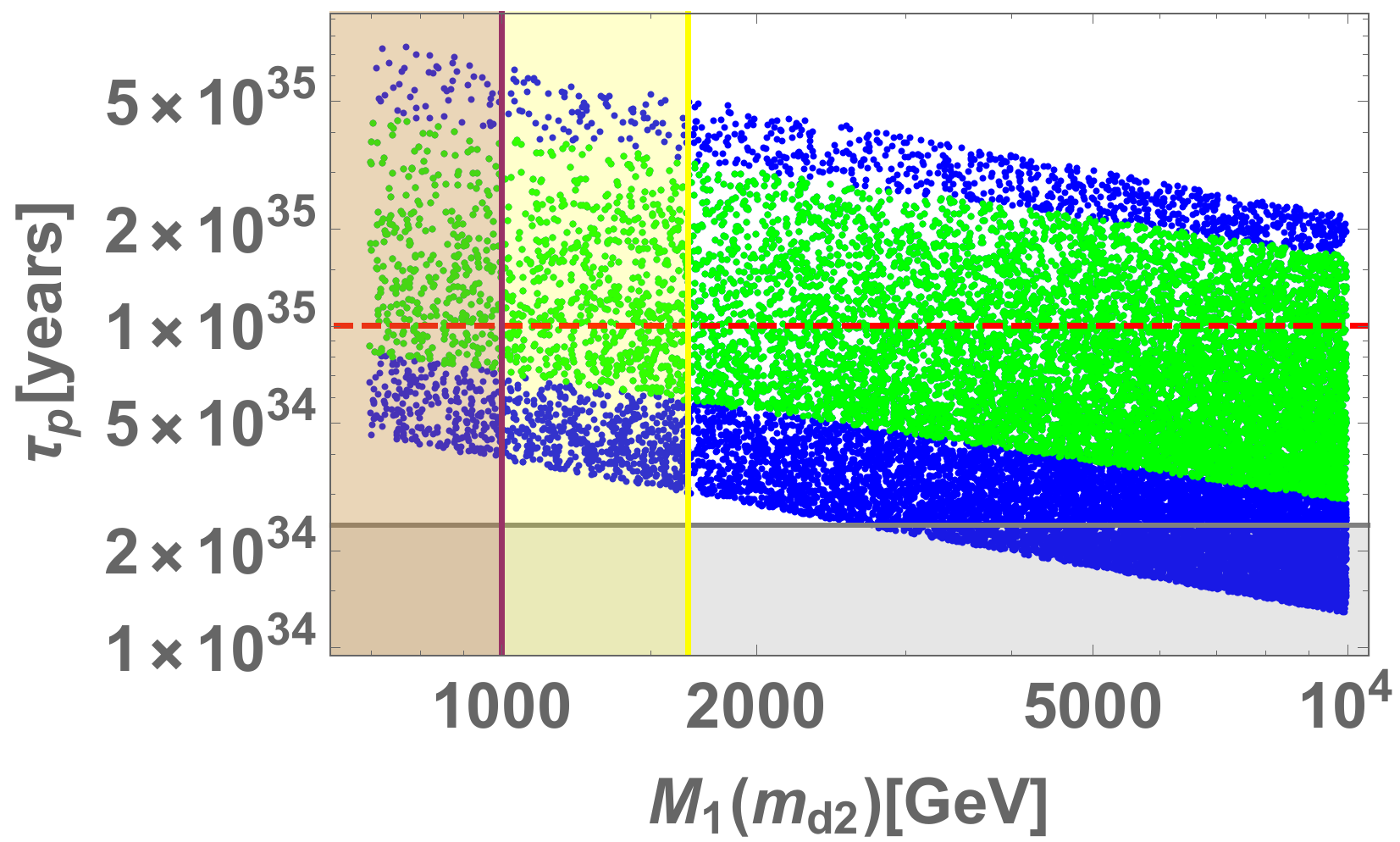}
    \subcaption{}
    \label{M1md2-tau}
    \end{minipage} & 
    \begin{minipage}[t]{0.45\hsize}
        \centering
        \includegraphics[keepaspectratio,scale=0.45]{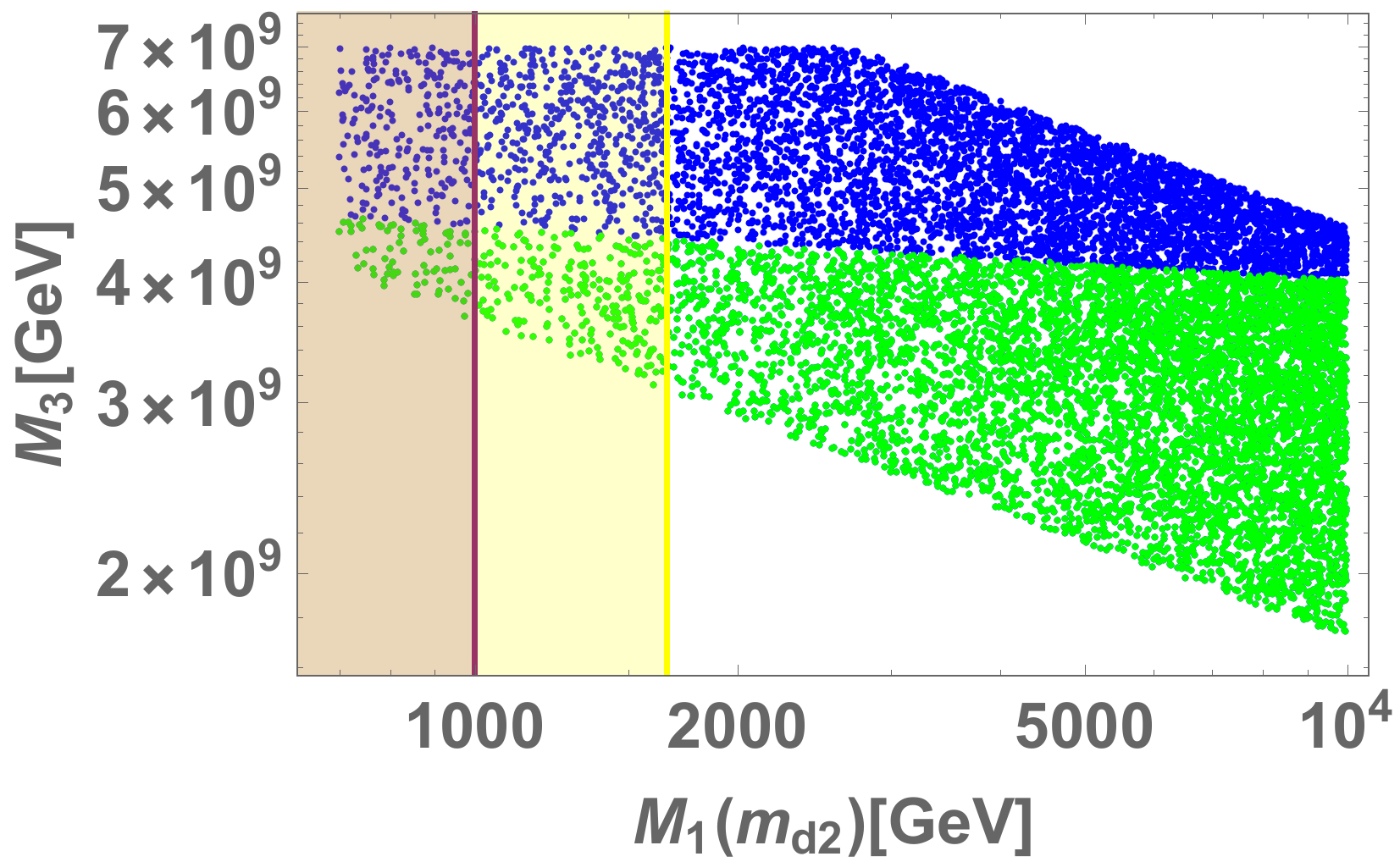}
    \subcaption{}
    \label{M1-M3}
    \end{minipage}\\
    \begin{minipage}[t]{0.45\hsize}
    \centering
    \includegraphics[keepaspectratio,scale=0.45]{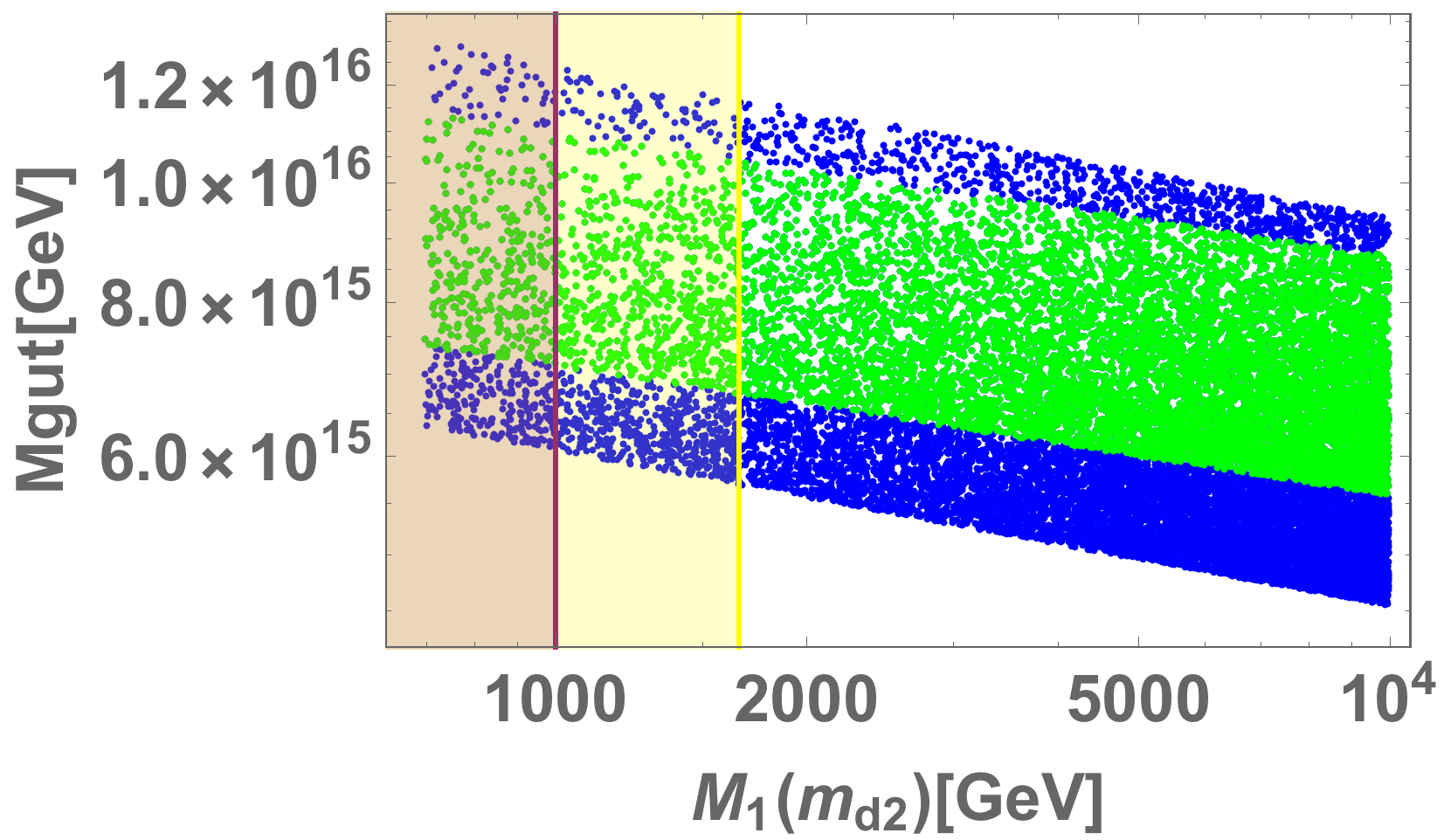}
    \subcaption{}
    \label{M1md2-Mgut}
    \end{minipage} & 
    \begin{minipage}[t]{0.45\hsize}
        \centering
        \includegraphics[keepaspectratio,scale=0.45]{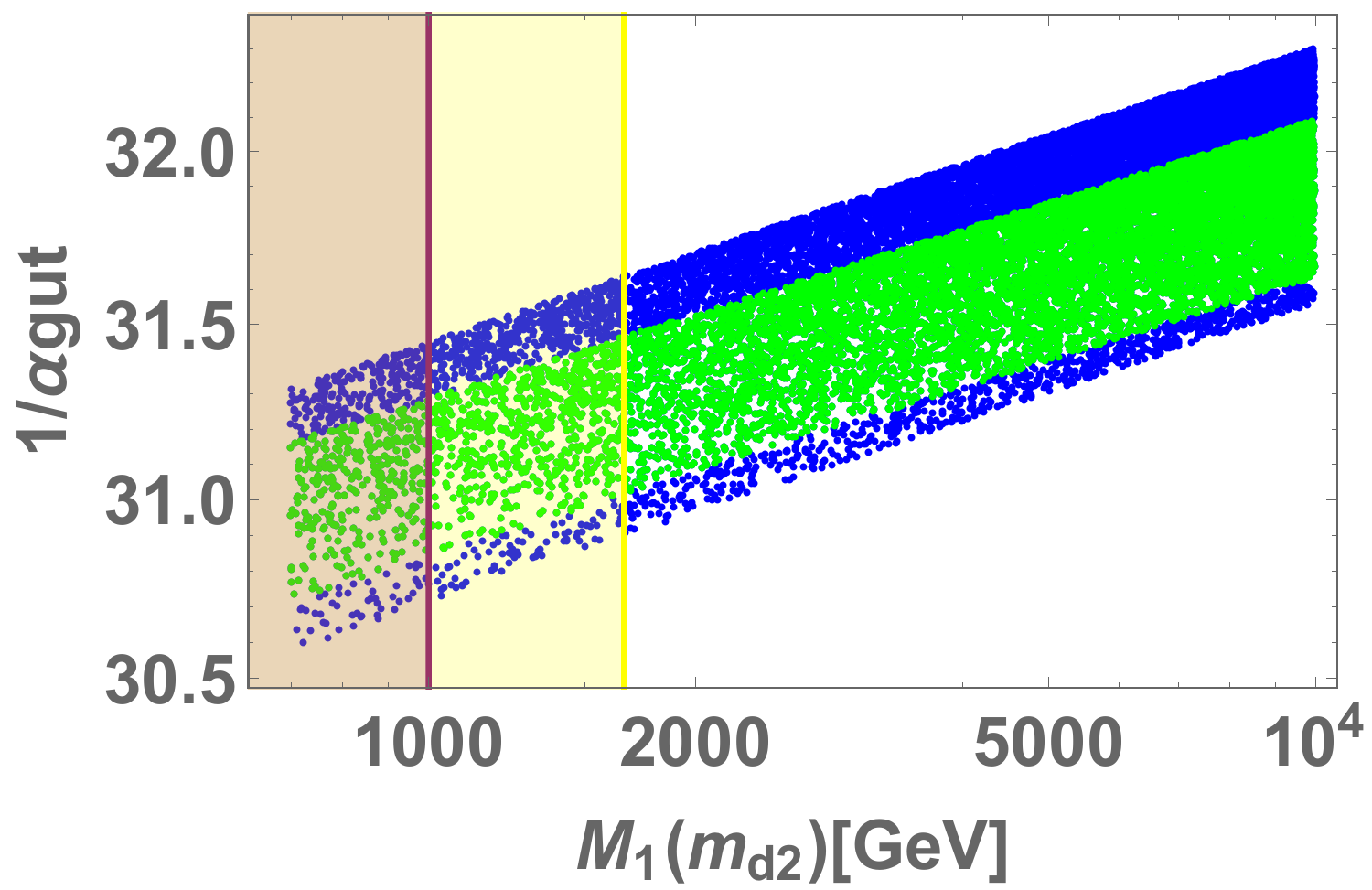}
    \subcaption{}
    \label{M1md2-alpha}
    \end{minipage}\\
    \end{tabular}
    \caption{In the above four figures, the horizontal axis is the scalar $\phi_1$ mass ($M_1$). The vertical axis is \eqref{M1md2-tau}: proton lifetime ($\tau_p$), \eqref{M1-M3}: scalar $\phi_3$ mass ($M_3$), \eqref{M1md2-Mgut}: the scale of unifying the SM gauge couplings ($M_\mathrm{GUT}$), and \eqref{M1md2-alpha}: the inverse of the GUT coupling (1/$\alpha_\mathrm{GUT}$), respectively. The yellow shaded region shows the bound of the vector-like quark mass, $m_{d2}>1660$~GeV~\cite{CMS:2018dcw}, and the purple shaded region shows the bound of the color octet scalar mass, $M_1>1$~TeV~\cite{Hayreter:2017wra, Miralles:2019uzg}. The green~(blue) shaded region denotes the range of unifying the SM gauge couplings with an accuracy of 1~\%~(3~\%) or less. In figure \eqref{M1md2-tau}, the gray shaded region shows the excluded one from the Super-Kamiokande experiment, $\tau_p(p\to\pi^0{e^+})>2.4\times10^{34}$~years~\cite{Super-Kamiokande:2020wjk}, and the red dashed line depicts the expected proton lifetime limit from the Hyper-Kamiokande experiment, $\tau_p(p\to\pi^0{e^+})<1.0\times10^{35}$~years~\cite{Dealtry:2019ldr}.}
    \label{M1comparison}
\end{figure}

Next, we evaluate the unifying of the SM gauge couplings and proton lifetime including the contribution of the new fermion and scalar masses.
The SM gauge couplings for $\mathrm{U(1)_Y},~\mathrm{SU(2)_L},$ and $\mathrm{SU(3)_C}$ are defined as $\alpha_i=g_i/4\pi$~$(i=1$-$3)$, respectively.
In our analysis, we assume $m_{d2}$ and $M_1$ are the same scale for simplicity.
We show the relations between the scalar $\phi_1$ mass ($M_1$) and proton lifetime ($\tau_p$), the scalar $\phi_3$ mass ($M_3$), the scale of unifying the SM gauge couplings ($M_\mathrm{GUT}$), and the inverse of the GUT coupling (1/$\alpha_\mathrm{GUT}$) in Figure~\ref{M1comparison}.
The yellow shaded region shows the bound of the 
vector-like quark mass, $m_{d2}>1660$~GeV~\cite{CMS:2018dcw}, and the purple shaded region shows the bound of the color octet scalar mass, $M_1>1$~TeV~\cite{Hayreter:2017wra, Miralles:2019uzg}. 
The green~(blue) shaded region denotes the range of unifying the SM gauge couplings with an accuracy of 1~\%~(3~\%) or less.
We define the accuracy of the unification as a percentage difference between the energy scale of unifying the SM gauge couplings $\alpha_1,~\alpha_2$ and $\alpha_2,~\alpha_3$.
In figure \ref{M1md2-tau}, the gray shaded region shows the excluded one from the Super-Kamiokande experiment, $\tau_p(p\to\pi^0{e^+})>2.4\times10^{34}$~years~\cite{Super-Kamiokande:2020wjk}, and the red dashed line depicts the expected proton lifetime limit from the Hyper-Kamiokande experiment, $\tau_p(p\to\pi^0{e^+})<1.0\times10^{35}$~years~\cite{Dealtry:2019ldr}.
If experiments about the vector-like quarks or the color octet scalar are more precise, we find that the region within an accuracy of 1~\% can be tested by the Hyper-Kamiokande experiment is more extended.
In figure \ref{M1-M3}, we denote the region that is allowed by the Super-Kamiokande experiment and can be tested by the Hyper-Kamiokande one.

\begin{figure}[htbp]
\begin{minipage}[b]{0.5\linewidth}
    \centering
    \includegraphics[keepaspectratio,scale=0.45]{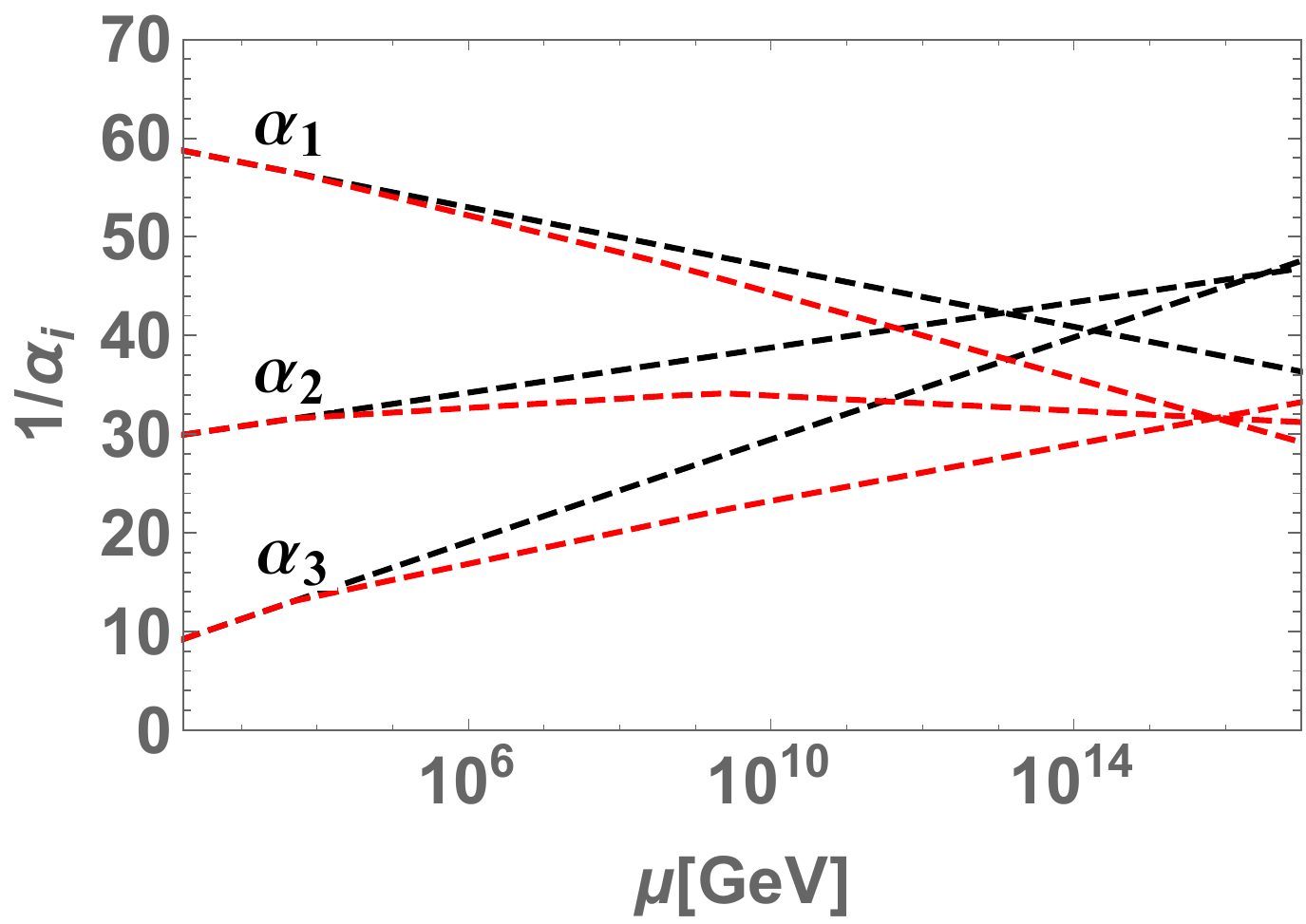}
    \subcaption{}
    \label{fig:unification}
\end{minipage}  
\begin{minipage}[b]{0.5\linewidth}
    \centering
    \includegraphics[keepaspectratio,scale=0.45]{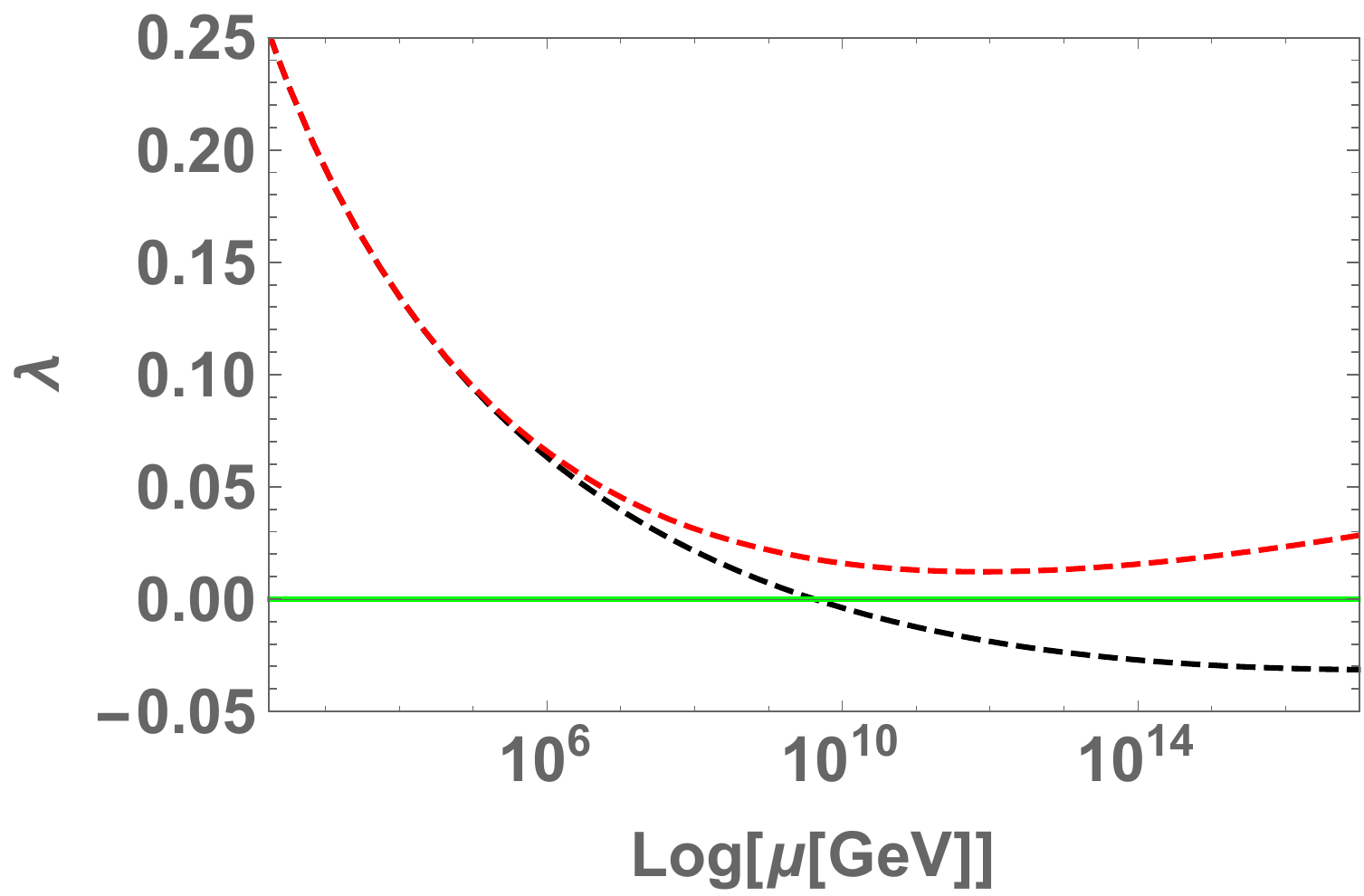}
    \subcaption{}
    \label{fig:lambda}
\end{minipage}  
\caption{The figure~\eqref{fig:unification}: the running of the SM gauge couplings. The figure~\eqref{fig:lambda}: the running of the SM Higgs quartic coupling. In both figures, the black dashed line depicts the RGE solutions of only the SM particles and the red dashed line depicts these including the new particle contributions. In figure~\eqref{fig:unification}, the SM gauge couplings unify successfully at the $M_\mathrm{GUT}\approx{7.28}\times10^{15}$~GeV and $\alpha_\mathrm{GUT}=\alpha_1=\alpha_2=\alpha_3\approx{1/31.7}$. In figure~\eqref{fig:lambda}, the horizontal green line depicts $\lambda=0$. Thanks to the new particle contributions, the SM Higgs quartic coupling is the positive value at all energy scales and the SM Higgs potential is stabilized.}
\label{SMcoupling}
\end{figure}

We set the benchmark that $m_{d2}=M_1=5$~TeV, $m_{e2}=3.68\times10^{8}$~GeV, and $M_3=2.5\times10^9$~GeV.
By using this benchmark, we estimate $M_\mathrm{GUT}$ and proton lifetime $\tau_p$.
We show the unification of the SM gauge couplings and the running of the SM Higgs quartic coupling in Figure~\ref{SMcoupling}.
The black dashed line depicts the RGE solutions of only the SM particles and the red dashed line depicts these including the new particle contributions.
In the case that includes the new particle contributions, the SM gauge couplings unify successfully at the $M_\mathrm{GUT}\approx{7.28}\times10^{15}$~GeV and $\alpha_\mathrm{GUT}=\alpha_1=\alpha_2=\alpha_3\approx{1/31.7}$.
The SM Higgs quartic coupling is the positive value at all energy scales including the new particle contributions and the SM Higgs potential is stabilized.
Also, by using these values, we estimate the proton lifetime approximately~\cite{Nath:2006ut}, 
\begin{equation}
    \tau_p(p\to\pi^0{e^+})\approx{\frac{1}{\alpha^2_\mathrm{GUT}}}\frac{M^4_\mathrm{GUT}}{m^5_p}\approx{8.07\times10^{34}}~\mathrm{years},
\end{equation}
where $m_p=0.938$~GeV is the proton mass~\cite{ParticleDataGroup:2022pth}.
The current proton lifetime limit from the Super-Kamiokande experiment is $\tau_p(p\to\pi^0{e^+})>2.4\times10^{34}$~years~\cite{Super-Kamiokande:2020wjk} and  the expected proton lifetime limit from the Hyper-Kamiokande experiment is $\tau_p(p\to\pi^0{e^+})<1.0\times10^{35}$~years~\cite{Dealtry:2019ldr}.
Therefore, it is consistent with the current proton lifetime limit from the Super-Kamiokande experiment and can be tested by the Hyper-Kamiokande experiment.
The proton decay can be mediated by the color triplet scalar field contained in the $\mathbf{5}$ representation Higgs and the $\mathbf{45}$ representation Higgs.
The Super-Kamiokande experiment excludes the colored scalar mass that is lighter than $\mathcal{O}(10^{11})$~GeV~\cite{Nath:2006ut}.
Considering the cross terms for the $H$ and $\Sigma$, for example, $H^\dagger{H}\Tr[\Sigma^\dagger{\Sigma}]$, the color triplet scalar field contained in the $\mathbf{5}$ representation Higgs is greater than $\mathcal{O}(10^{11})$~GeV.
Since the couplings which induce the proton decay by mediating the $\phi_3$ are absent, the $\phi_3$ cannot induce the proton decay and thus it can be light.

\newpage
\section{Estimate of heavy neutrino masses}
\label{sec:neutrino_mass}
In this section, we obtain the mass relations among the neutrinos and estimate the heavy neutrino masses.
First, we analyze the neutrino mass eigenvalues by using the $9\times9$ mass matrix $\mathcal{M}_\nu$ in Eq.~\eqref{eq:mass_matrices}:
\begin{equation}
\label{eq:mass_matrices1}
    \mathcal{M}_\nu=
    \begin{pmatrix}
        0&\Tilde{M}_D&M_D\\
        \Tilde{M}_D^T&0&0\\
        M_D^T&0&M_R\\
    \end{pmatrix},
\end{equation}
where the $3\times3$ mass matrices are defined as
\begin{align}
\label{eq:33massmatrix}
    \Tilde{M}_D&=\Tilde{Y}_5^{ij}\frac{3v_\Sigma}{4\sqrt{15}},\nonumber\\
    M_D&=Y_D^{ij}\frac{v_H}{\sqrt{2}},\\
    M_R&=Y_M^{i}\frac{v_\Phi}{\sqrt{2}}.\nonumber
\end{align}
In our analysis of neutrino masses, we also assume the diagonal matrix for simplicity.
By using the type-I seesaw mechanism~\cite{Minkowski:1977sc, Yanagida, Gell-Mann:1979vob, Mohapatra:1979ia, Schechter:1980gr}, the Eq.~\eqref{eq:mass_matrices1} is rewritten as
\begin{equation}
\label{eq:neutrinomassmatrix}
    \mathcal{M}_\nu=
    \begin{pmatrix}
        M_\nu&\Tilde{M}_D&0\\
        \Tilde{M}_D^T&0&0\\
        0&0&\sim{M_R}\\
    \end{pmatrix}.
\end{equation}
Then, we diagonalize $\mathcal{M}_\nu$ in Eq.~\eqref{eq:neutrinomassmatrix} and get the neutrino mass eigenvalues,
\begin{equation}
\label{eq:neutrinomassdiagonalmatrix}
    \mathcal{M}_\nu=
    \begin{pmatrix}
        m_\nu&0&0\\
        0&\Tilde{m}_\nu&0\\
        0&0&\sim{M_R}\\
    \end{pmatrix},
\end{equation}
where $m_\nu$ is the left-handed Majorana neutrino mass matrix, $\Tilde{m}_\nu$ is the mirror neutrino mass matrix, and $M_R$ is the right-handed Majorana neutrino mass matrix.
The left-handed Majorana and the mirror neutrino mass eigenvalues are related to $\Tilde{Y}_{5}$ matrix diagonal elements in Eq.~\eqref{eq:33massmatrix}.
The down-type quark mass eigenvalues in Eq.~\eqref{eq:downtypemasseigenvalues} also include the $\Tilde{Y}_{5}$ matrix diagonal elements.
Since the $\Tilde{Y}_{5}$ diagonal elements include the up- and down-type quark mass eigenvalues, the left-handed Majorana and the mirror neutrino mass eigenvalues are represented in terms of the up- and down-type quark mass eigenvalues.
We show the left-handed Majorana and the mirror neutrino mass eigenvalues $m_i~(i=1$-$6)$ just below:
\begin{equation}
\label{eq:neutrinomass}
    \begin{aligned}
        m_1&=\frac{\big(v_H{Y_D^{11}}\big)^2-\sqrt{\Big[\big({v_H}Y_D^{11}\big)^4+\frac{72m_{d1}^2m_{d2}^2}{m_{u1}m_{u2}}\big({v_\Phi}{Y_M^1}\big)^2\Big]}}{2\sqrt{2}v_\Phi{Y_M^1}},\\
       m_2&=\frac{\big(v_H{Y_D^{11}}\big)^2+\sqrt{\Big[\big({v_H}Y_D^{11}\big)^4+\frac{72m_{d1}^2m_{d2}^2}{m_{u1}m_{u2}}\big({v_\Phi}{Y_M^1}\big)^2\Big]}}{2\sqrt{2}v_\Phi{Y_M^1}},\\
        m_3&=\frac{\big(v_H{Y_D^{22}}\big)^2-\sqrt{\Big[\big({v_H}Y_D^{22}\big)^4+\frac{72m_{d3}^2m_{d4}^2}{m_{u3}m_{u4}}\big({v_\Phi}{Y_M^2}\big)^2\Big]}}{2\sqrt{2}v_\Phi{Y_M^2}},\\
        m_4&=\frac{\big(v_H{Y_D^{22}}\big)^2+\sqrt{\Big[\big({v_H}Y_D^{22}\big)^4+\frac{72m_{d3}^2m_{d4}^2}{m_{u3}m_{u4}}\big({v_\Phi}{Y_M^2}\big)^2\Big]}}{2\sqrt{2}v_\Phi{Y_M^2}},\\
        m_5&=\frac{\big(v_H{Y_D^{33}}\big)^2-\sqrt{\Big[\big({v_H}Y_D^{33}\big)^4+\frac{72m_{d5}^2m_{d6}^2}{m_{u5}m_{u6}}\big({v_\Phi}{Y_M^3}\big)^2\Big]}}{2\sqrt{2}v_\Phi{Y_M^3}},\\
        m_6&=\frac{\big(v_H{Y_D^{33}}\big)^2+\sqrt{\Big[\big({v_H}Y_D^{33}\big)^4+\frac{72m_{d5}^2m_{d6}^2}{m_{u5}m_{u6}}\big({v_\Phi}{Y_M^3}\big)^2\Big]}}{2\sqrt{2}v_\Phi{Y_M^3}}.\\
    \end{aligned}
\end{equation}
By using these eigenvalues, we obtain the mass relations among the neutrinos,
\begin{equation}
\label{eq:neutrinorelation}
    \begin{aligned}
        m_1m_2&=\frac{9m_{d1}^2m_{d2}^2}{m_{u1}m_{u2}},\\
        m_3m_4&=\frac{9m_{d3}^2m_{d4}^2}{m_{u3}m_{u4}},\\
        m_5m_6&=\frac{9m_{d5}^2m_{d6}^2}{m_{u5}m_{u6}}.\\
    \end{aligned}
\end{equation}
Now, we assume that $m_1$, $m_3$, and $m_5$ are the left-handed Majorana neutrino masses.
Since the neutrino masses are related to the up- and down-type quark masses, we can get the product of two neutrino masses.
Then, we estimate the heavy neutrino masses by using Eq.~\eqref{eq:neutrinorelation}.
There are two patterns about the heavy neutrino masses.
\newpage
\begin{itemize}
    \item [1]. The $m_{u3}$ is the SM particle, $m_{u4}$ is $\mathcal{O}(M_{\mathrm{GUT}})$, $m_{d3}$ is the SM particle, and $m_{d4}$ is $\mathcal{O}(M_{\mathrm{GUT}})$.

    In this case, the neutrino mass relation is 
    \begin{equation}
        m_3m_4=\mathcal{O}(M_{\mathrm{GUT}}).
    \end{equation}
    Then, the heavy neutrino mass is the GUT scale.
    The mass relation between $m_5$ and $m_6$ is the same result.

    \item[2]. The $m_{u1}$ is the SM particle, $m_{u2}$ is $\mathcal{O}(M_{\mathrm{GUT}})$, $m_{d1}$ is bottom quark, and $m_{d2}$ is $\mathcal{O}(10^3)$~GeV.

    By using the benchmark that $m_{d1}=4.18$~GeV, $m_{d2}=5$~TeV, and $m_{u2}=M_{\mathrm{GUT}}={7.28}\times10^{15}$~GeV, the neutrino mass relation is
    \begin{equation}
    \label{eq:numassrelation}
        m_1m_2
        \approx\frac{5.40\times{10^{-7}}}{m_{u1}}~\mathrm{(GeV)}^2.\\
    \end{equation}
\end{itemize}
In Eq.~\eqref{eq:numassrelation}, $m_{u1}$ is three patterns such as the up quark, charm quark, and top quark.
Then, we get three neutrino mass relations:
\begin{equation}
\label{eq:eachnumassrelation}     m_1m_2=2.50\times{10}^{14}~\mathrm{(eV)^2},\quad{m_1m_2}=4.25\times{10}^{11}~\mathrm{(eV)^2},\quad{m_1m_2}=3.13\times{10}^{9}~\mathrm{(eV)^2}.
\end{equation}
By using Eq.~\eqref{eq:eachnumassrelation}, we estimate the heavy neutrino masses.
There are two neutrino mass hierarchies, normal hierarchy (NH) and inverted hierarchy (IH).
We can derive the upper bound of the lightest neutrino masses from the experimental results of the neutrino mass squared differences~\cite{Esteban:2020cvm, Gonzalez-Garcia:2021dve} and the sum of neutrino masses~\cite{Planck:2018vyg}.
In the NH, the upper bound of the lightest neutrino mass is 0.03~eV.
Then, the lower bound of the heavy neutrino masses is obtained just below:
\begin{equation}
\label{eq:normalneutrino}
m_{\mathrm{heavy}}>8.33\times{10}^6~\mathrm{GeV},\quad{m_{\mathrm{heavy}}}>1.42\times{10}^4~\mathrm{GeV},\quad{m_{\mathrm{heavy}}}>104.34~\mathrm{GeV}.
\end{equation}
On the other hand, in the IH, the upper bound of the lightest neutrino mass is 0.016~eV.
Then, the lower bound of the heavy neutrino masses is obtained just below:
\begin{equation}
\label{eq:invertedneutrino}
m_{\mathrm{heavy}}>1.56\times{10}^7~\mathrm{GeV},\quad{m_{\mathrm{heavy}}}>2.66\times{10}^4~\mathrm{GeV},\quad{m_{\mathrm{heavy}}}>195.63~\mathrm{GeV}.
\end{equation}
Therefore, the lower bound of the heavy neutrino masses in the top quark case is $\mathcal{O}(10^2)$~GeV and it is testable for the future experiment about the sterile neutrinos\cite{Alekhin:2015byh}.
\section{Summary and Discussions}
\label{sec:Summary}

We have proposed an SU(5)$\times $U(1)$_\text{X}\times$U(1)$_\text{PQ}$ model without SUSY. 
The $\mathrm{U(1)_{X}}$ gauge symmetry is the generalization of the $\mathrm{U(1)_{B-L}}$ gauge symmetry.
The $\mathrm{U(1)_{PQ}}$ symmetry is the global PQ symmetry.
The PQ symmetry can solve the strong CP problem and the pseudo-Nambu Goldstone boson from the PQ symmetry breaking called the axion is the DM candidate.
In previous work, they added one family of the same representation for the SM particles and one mirror family.
The mirror family is the conjugate of the representation for the SM particles.
We have introduced only three mirror families in order to unify the SM gauge couplings and avoid the restriction of proton lifetime. 
The SM matter fields are unified into $\mathbf{\Bar{5}}$+$\mathbf{10}$ representation of SU(5) and the mirror family into $\mathbf{5}$+$\mathbf{\overline{10}}$ representation.
We also have introduced three gauge singlet right-handed Majorana neutrinos to derive the left-handed Majorana neutrino masses through the type-I seesaw mechanism and cancel all the U(1)$_\text{X}$ related anomalies.
In order to obtain the difference of the masses for the down-type quarks and charged leptons, we have introduced the $\mathbf{45}$ representation Higgs in addition to the $\mathbf{5}$ representation and $\mathbf{24}$ one.
In order to occur the inflation and break the U(1)$_\text{X}$ gauge symmetry, the gauge singlet scalar field is introduced.
Then, we have discussed the mass relations between the SM and mirror particles and identified the mass scales of the mirror particles.
We have set the benchmark for the mass scales of the mirror particles in our analysis.
Since the new particles exist in the intermediate energy scale and contribute to the RGE, the SM gauge couplings unify successfully at $M_\mathrm{GUT}\approx{7.28}\times10^{15}$~GeV and $\alpha_\mathrm{GUT}=\alpha_1=\alpha_2=\alpha_3\approx{1/31.7}$. 
Our model expects the proton lifetime as $\tau_p(p\to\pi^0{e^+})\approx{8.07\times10^{34}}~\mathrm{years}$ and can be tested by the future proton decay search, e.g., the Hyper-Kamiokande experiment expected as $\tau_p(p\to\pi^0{e^+})<1.0\times10^{35}$ years.
Also, by using the mass relations between the active and mirror neutrinos, we have estimated the lower bound of the heavy neutrino masses.
The lowest bound of the heavy neutrino masses is ${m_{\mathrm{heavy}}}>104.34~\mathrm{GeV}$ and it is testable for the future experiment about the sterile neutrinos. 

In our model, the heavy neutrinos have the potential to be the candidate for DM in addition to the axion.
Since the SU(5) symmetry breaking and the $\mathrm{U(1)_{PQ}}$ symmetry breaking occur at the same energy scale, the scenarios of inflation are restricted by the axion domain wall, axion DM isocurvature~\cite{Kawasaki:2013ae, Okada:2020cvq}, and SU(5) monopole problems~\cite{tHooft1974, Polyakov:1974ek, Langacker:1980kd, Preskill:1984gd}.
Then, we will discuss the scenarios of dark matter and inflation in future work.

\vspace{1cm}
\noindent
{\large \bf Acknowledgement}
\vspace{1mm}

We would like to thank Prof. N. Okada for useful discussions and comments.

\appendix
\section*{Appendix}
\section{The relevant representations in SU(5)}
\label{sec:Representations}
We show the unification of the SM particles into the SU(5) representations.
The SM matter fields are unified into 
$\mathbf{\bar 5}+\mathbf{10}$ just below:
\begin{equation}
\begin{aligned}
    \psi_{\bar{5}}&=
    \begin{pmatrix}
    d^c_1\\
    d^c_2\\
    d^c_3\\
    e\\
    -\nu\\
    \end{pmatrix}_L,\quad
    \psi_{10}=\frac{1}{\sqrt{2}}
    \begin{pmatrix}
    0&u_3^c&-u_2^c&-u^1&-d^1\\
    -u_3^c&0&u_1^c&-u^2&-d^2\\
    u_2^c&-u_1^c&0&-u^3&-d^3\\
    u^1&u^2&u^3&0&-e^+\\
    d^1&d^2&d^3&e^+&0\\
   \end{pmatrix}_L.
\end{aligned}
\end{equation}
Here, $``c"$ means charge conjugation, and the SM gauge bosons are unified into 
$\mathbf{24}$ just below:
\begin{equation}
    \mathbf{24}=
    \begin{pmatrix}
        G_\mu-\frac{1}{\sqrt{15}}B_\mu&V^\dagger_\mu\\
        V_\mu&W_\mu+\frac{3}{2\sqrt{15}}B_\mu\\
    \end{pmatrix},
\end{equation}
with quantum numbers as
\begin{equation}
G_\mu\sim(8,1,0),\quad{W_\mu}\sim(1,3,0),\quad{B_\mu}\sim(1,1,0),\quad{V}^\mu\sim(3,2,-5/6).
\end{equation}
On the other hand, the $\mathbf{5}$ representation Higgs $H$ and $\mathbf{24}$ representation Higgs $\Sigma$ are unified as follows:
\begin{equation}
    H=
        \begin{pmatrix}
            H_1\\
            H_2\\
            H_3\\
            H^+\\
            H^0\\
        \end{pmatrix},\quad
        \Sigma=
        \begin{pmatrix}
            \Sigma_8-\frac{2}{\sqrt{30}}\Sigma_0&\Sigma_{(\bar{3},2)}\\
           \Sigma_{(3,2)}&\Sigma_3+\frac{3}{\sqrt{30}}\Sigma_0
        \end{pmatrix},\\
\end{equation}    
where $H_1$, $H_2$, and $H_3$ are colored Higgs and $H^+$ and $H^0$ are the SM Higgs doublet.
The component of the $\mathbf{24}$ representation Higgs have the quantum numbers just below:
\begin{equation}
\Sigma_8\sim(8,1,0),\quad\Sigma_3\sim(1,3,0),\quad\Sigma_0\sim(1,1,0),\quad\Sigma_{(3,2)}\sim(3,2,-\frac{5}{6}),\quad\Sigma_{(\bar{3},2)}\sim(\bar{3},2,\frac{5}{6}),\\
\end{equation}
and the $\mathbf{45}$ representation Higgs $\chi$ has the following components,
\begin{align}
    \chi\sim\Phi_1(8,2,\frac{1}{2})\oplus\Phi_2(\bar{6},1,-\frac{1}{3})&\oplus\Phi_3(3,3,-\frac{1}{3})\nonumber\\
    &\oplus\Phi_4(\bar{3},2,-\frac{7}{6})
\oplus\Phi_5(3,1,-\frac{1}{3})\oplus\Phi_6(\bar{3},1,\frac{4}{3})\oplus{H_2}(1,2,\frac{1}{2}). 
\end{align}

\section{The RGE and beta coefficients}
\label{sec:beta_coefficients}
We show the RGE form,
\begin{equation}
    \mu\frac{dg_i}{d\mu}=g_i^3[\beta_{gi}(\mathrm{SM})+\beta_{gi}(\mathrm{NEW})],
\end{equation}
where $g_i$ $(i=1\mathrm{-}3)$ are the SM gauge couplings, $\beta_{gi}(\mathrm{SM})$ are the contributions of the SM particles, and $\beta_{gi}(\mathrm{NEW})$ are the new particle contributions, respectively.
In our analysis, we consider the contributions of the SM particles at 2-loop level~\cite{Machacek:1983tz, Machacek:1983fi, Machacek:1984zw} and the new particle contributions at 1-loop level such as
\begin{equation}
    \beta_{gi}(\mathrm{NEW})=\frac{1}{16\pi^2}[b_i\times\theta(\mu-M)],
\end{equation}
where $M$ is the mass of each field and the beta coefficients $b_i$ for each field are listed in Table~\ref{tab:betacoefficients}.
Here, $\theta(\mu-M)$ is a step function and we add the new particle contributions as the step function for each particle.
\begin{table}[h]
    \centering
    \begin{tabular}{|c|c|c|c|}
    \hline
         Fields&$b_1$&$b_2$&$b_3$\\
         \hline\hline
         $\Phi_1$&4/5&4/3&2\\
         \hline
         $\Phi_2$&2/15&0&5/6\\
         \hline
         $\Phi_3$&1/5&2&1/2\\
         \hline
         $\Phi_4$&49/30&1/2&1/3\\
         \hline
         $\Phi_5$&1/15&0&1/6\\
         \hline
         $\Phi_6$&16/15&0&1/6\\
         \hline
         $H_2$&1/10&1/6&0\\
         \hline
         $m_{ui}$&17/30&1/2&2/3\\
         \hline
         $m_{di}$&1/6&1/2&2/3\\
         \hline
         $m_{ei}$&3/5&1/3&0\\
         \hline
    \end{tabular}
    \caption{The beta coefficients for each field.}
    \label{tab:betacoefficients}
\end{table}

\end{document}